\documentclass[twocolumn,prl,showpacs,superscriptaddress]{revtex4}
\usepackage{amsmath}
\usepackage{graphicx}
\usepackage{amssymb}
\usepackage{dcolumn}
\usepackage{mathrsfs}
\usepackage{bm}

\begin{document}

\title{Subnatural-Linewidth Polarization-Entangled Photon Pairs with Controllable Temporal Length}

\author{Kaiyu Liao}
 \affiliation{Laboratory of Quantum
Engineering and Quantum Materials,School of Physics and
Telecommunication Engineering, South China Normal University,
Guangzhou 510006, China}

\author{Hui Yan}
\email{yanhui@scnu.edu.cn}
 \affiliation{Laboratory of Quantum
Engineering and Quantum Materials,School of Physics and
Telecommunication Engineering, South China Normal University,
Guangzhou 510006, China}

\author{Junyu He}
 \affiliation{Laboratory of Quantum
Engineering and Quantum Materials,School of Physics and
Telecommunication Engineering, South China Normal University,
Guangzhou 510006, China}

\author{Shengwang Du}
\affiliation{Department of Physics, The Hong Kong University of
Science and Technology, Clear Water Bay, Kowloon, Hong Kong,
China}

\author{Zhi-Ming Zhang}
 \affiliation{Laboratory of Quantum
Engineering and Quantum Materials,School of Information and
Photoelectronic Science and Engineering, South China Normal
University, Guangzhou 510006, China}

\author{Shi-Liang Zhu}
\email{slzhunju@163.com} \affiliation{National Laboratory of Solid
State Microstructures and School of Physics, Nanjing University,
Nanjing 210093, China}
 \affiliation{Laboratory of
Quantum Engineering and Quantum Materials,School of Physics and
Telecommunication Engineering, South China Normal University,
Guangzhou 510006, China}


\begin{abstract}
We demonstrate an efficient experimental scheme for producing
polarization-entangled photon pairs from spontaneous four-wave
mixing (SFWM) in a laser-cooled $^{85}$Rb atomic ensemble, with a
bandwidth (as low as 0.8 MHz) much narrower than the rubidium
atomic natural linewidth. By stabilizing the relative phase
between the two SFWM paths in a Mach-Zehnder interferometer
configuration, we are able to produce all four Bell states. These
subnatural-linewidth photon pairs with polarization entanglement
are ideal quantum information carriers for connecting remote
atomic quantum nodes via efficient light-matter interaction in a
photon-atom quantum network.
\end{abstract}

 \maketitle

The connectivity of a long-distance photon-atom quantum network
strongly depends on efficient interactions between flying photonic
quantum bits and local long-lived atomic matter nodes \cite{Duan,
KimbleNature2008}. Such efficient quantum interfaces, which
convert quantum states (such as time-frequency waveform and
polarizations) between photons  and atoms, require the photons to
have a bandwidth sufficiently narrower than the natural linewidth
of related atomic transitions (such as 6 MHz for rubidium D1 and
D2 lines). As a standard method for producing entangled photons,
spontaneous parametric down conversion (SPDC) in a nonlinear
crystal usually has a wide bandwidth (larger than terahertz) and
very short coherence time (less than picosecond). Many efforts
have been investigated in the past more than one decade to narrow
down the SPDC photon bandwidth using optical cavities
\cite{OuPRL1999, WangPRA2004, PanPRL2008, 1, 2, 3}. However, the
bandwidth of SPDC polarization-entangled photon pairs is still
wider than most atomic transitions and leads to a very low
efficiency of storing these polarization states in a quantum
memory \cite{PanPRL2008, PanNatPhotonics2011}.

Our motivation was stimulated by the recent progress in generating
subnatural-linewidth biphotons by using continuous-wave
spontaneous four-wave mixing (SFWM) in a laser-cooled atomic
ensemble with electromagnetically induced transparency (EIT)
\cite{fwm2, review}. Photons produced from this method not only
have narrow bandwidth but also automatically match the atomic
transitions. The applications of these narrow-band photons include
the demonstration of a single-photon memory with a storage
efficiency of about 50\% \cite{memory1}, a single-photon precursor
\cite{DuPRL2011}, and coherent control of single-photon absorption
and reemission \cite{DuPRL2012}. However, while this method
provides a natural entanglement mechanism in the time-frequency
domain, it is extremely difficult to produce polarization
entanglement because of the polarization selectivity of EIT in a
nonpolarized atomic medium \cite{YuPRA2000}. It is possible to
generate the polarization entanglement by scarifying the EIT
effect, but the photon generation efficiency is low and the
bandwidth is not narrower than the atomic natural linewidth
\cite{yan}. The ``writing-reading'' technique with optical pumping
provides a solution to polarization entanglement but results in
reducing time-frequency entanglement \cite{KuzmichScience2004}.

In this Letter, we report our work on producing
subnatural-linewidth polarization-entangled photon pairs by using
the continuous-wave SFWM cold-atom EIT configuration. We
demonstrate that the polarization entanglement can be efficiently
produced by making use of a Mach-Zehnder interferometer in a
two-path SFWM setup while maintaining the EIT effect in
controlling the photon bandwidth. By tuning the phase difference
between the two SFWM paths and properly setting the driving laser
polarizations, we can generate all four Bell states. These photons
have a coherence time of up to 900 ns and an estimated bandwidth
of about 1 MHz that is much narrower than  the Rb atomic natural
linewidth (6 MHz). These subnatural-linewidth
polarization-entangled photon pairs are ideal flying qubits for
connecting remote atomic quantum nodes in a quantum network.

\begin{figure*}
\begin{center}
\includegraphics[width=18cm]{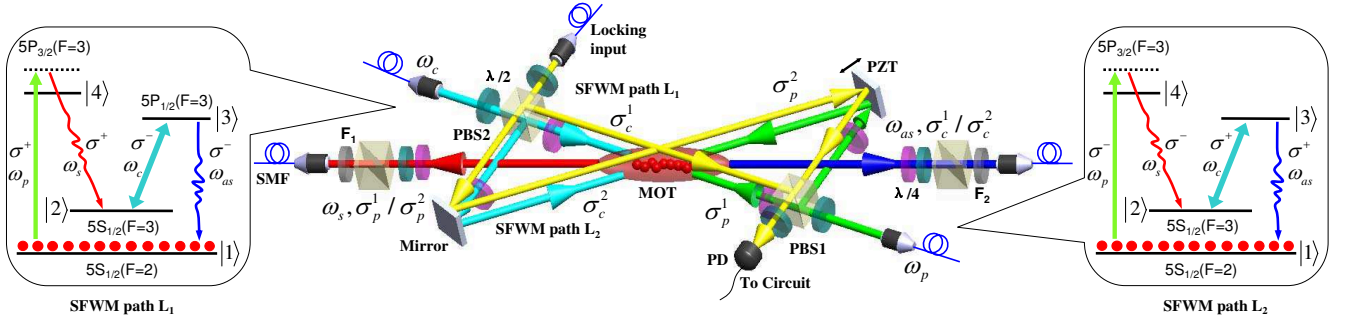}
\caption{\label{fig:1} (color online). Experimental setup for
producing subnatural-linewidth polarization-entangled photon
pairs. The polarization entanglement is created by the quantum
interference of the two spatially symmetric SFWM processes driven
by two counterpropagating pump-coupling beams ($L_1$ and $L_2$).
The  phase difference of the two SFWM paths is stabilized by
locking the reference laser in the Mach-Zehnder interferometer,
whose output is detected by a photodetector (PD). To avoid
interaction with the cold atoms, the reference beams are slightly
shifted away from the pump-coupling beams but pass through the
same optical components. The inserted energy level diagrams are
two possible SFWM channels for $L_{1}$ and $L_{2}$, respectively.
PZT: Piezoelectric transducer.}
\end{center}
\end{figure*}

Our experimental setup is illustrated in Fig. 1. We work with a
two-dimensional $^{85}$Rb magneto-optical trap (MOT) with a
longitudinal length of $L=1.7$ cm \cite{2DMOT}.  The experiment is
run periodically. In each cycle,  after 4.5 ms MOT time, the atoms
are prepared in the  ground level $\left| 1\right\rangle $ and
followed by a 0.5 ms SFWM biphoton generation window. Along the
longitudinal direction, the atoms have an optical depth of 32 in
the $ |1\rangle\rightarrow|3\rangle$ transition. The pump laser
(780 nm, $\omega_p$) is 80 MHz blue detuned from the transition
$\left| 1\right\rangle\to\left|4 \right\rangle$, and the coupling
laser (795 nm, $\omega_c$) is on resonance with the transition
$\left| 2 \right\rangle  \to \left| 3 \right\rangle $. The linear
polarized pump laser beam, with a $1/e^2$ diameter of 1.8 mm, is
equally split into two beams after a half-wave plate and the first
polarization beam splitter (PBS1). These two beams, with opposite
circular polarizations ($\sigma ^ + $ and $\sigma ^ -$) after two
quarter-wave plates, then intersect at the MOT with an angle of $
\pm 2.5^\circ$  to the longitudinal axis. Similarly, the two
coupling laser beams after PBS2 with opposite circular
polarizations overlap with the two pump beams from opposite
directions. In the presence of these two pairs of
counterpropagating pump-coupling beams, phase-matched Stokes
($\omega_s$) and anti-Stokes ($\omega_{as}$) paired photons are
produced along the longitudinal axis and coupled into two opposing
single-mode fibers (SMFs). In each SFWM path, the polarizations of
the Stokes and anti-Stokes photons follow those of the
corresponding pump and coupling field, respectively.  The two SMF
spatial modes are focused at the MOT center with a $1/e^2$
diameter of 0.4 mm. After two narrow-band filters (${F_1}$ and
${F_2}$, 0.5 GHz bandwidth), the photons are detected by two
single-photon counter modules (SPCM, Perkin Elmer SPCM-4Q4C) and
analyzed by a time-to-digital converter (Fast Comtec P7888) with a
time bin width of 1 ns. Two sets of quarter-wave plates, half-wave
plates, and PBSs are inserted for measuring polarization
correlation and quantum state tomography.

To obtain the polarization entanglement, we must stabilize the
phase difference between the two SFWM spatial paths. This is
achieved by injecting a reference laser beam (795 nm, 110 MHz blue
detuned from the transition $\left| 2 \right\rangle  \to \left| 3
\right\rangle $) from the second input of PBS2. The two reference
beams split after PBS2 are then recombined after PBS1 and detected
by a photodetector (a half-wave plate and a PBS are used to obtain
the interference), as shown in Fig. 1. This is a standard
Mach-Zehnder interferometer to the reference laser. Locking the
phase difference of the two arms of the Mach-Zehnder
interferometer with a feedback electronics stabilizes the phase of
the two SFWM paths (See Supplementary Material \cite{Supp}). To
avoid interaction with the cold atoms, the reference beams are
slightly shifted away from the pump-coupling beams but pass
through the same optical components.

\begin{figure}
\begin{center}
\includegraphics[width=8.5cm]{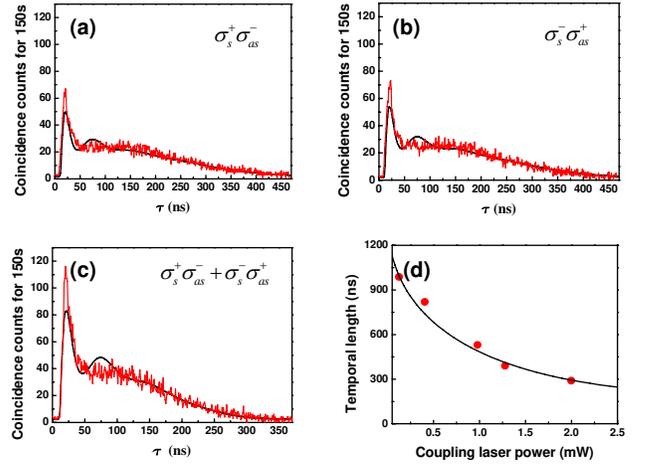}
\caption{\label{fig:2}(color online). (a)-(c) Two-photon
coincidence counts as a function of the relative time delay
between the Stokes and anti-Stokes photons. (a) Path 1:
$\sigma_p^1=\sigma^+$ and $\sigma_c^1=\sigma^-$; path 2: blocked.
(b) Path 1: blocked; path 2: $\sigma_p^2=\sigma^-$ and
$\sigma_c^2=\sigma^+$. (c) Both paths are present. (d) The
temporal length of the paired photons as a function of the
coupling laser power. All the black lines are theoretical plots.}
\end{center}
\end{figure}

\begin{figure}[tb]
\begin{center}
\includegraphics[width=7.5cm]{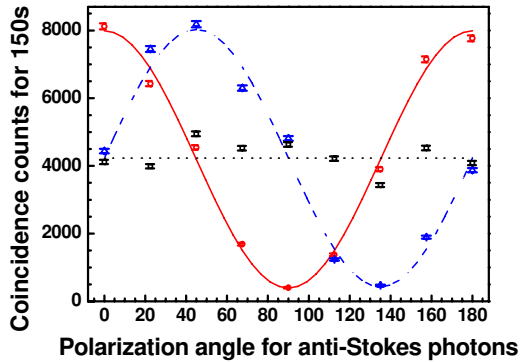}
\caption{\label{fig:3}(color online). Polarization correlation of
the paired Stokes and anti-Stokes photons at $\phi$=0. The
polarization angle for Stokes photons are set at $0^\circ$ (red
solid line and circle) and $-45^\circ$ (blue dashed line and
triangle). The black dotted line and squares are the data taken
without stabilizing $\phi$.}
\end{center}
\end{figure}

Following the perturbation theory \cite{review}, the produced
two-photon state can be described as (See Supplemental Material
\cite{Supp} for the derivation)
\begin{eqnarray}
\nonumber |\Psi\rangle && = L\int \kappa(\omega_{as})
\text{sinc}[\frac{\Delta
k(\omega_{as})L}{2}]|\omega_s=\omega_{p}+\omega_{c}-\omega_{as}\rangle\\
&& |\omega_{as}\rangle d\omega_{as}\otimes
\frac{1}{\sqrt{2}}(|\sigma_{p}^{1}\rangle_{s}|\sigma_{c}^{1}\rangle_{as}
+e^{i\phi}|\sigma_{p}^{2}\rangle_{s}|\sigma_{c}^{2}\rangle_{as} ),
\end{eqnarray}
which shows hyperentanglement in frequency (continuous) and
polarization (discrete). The upper indices (1 and 2) represent the
two SFWM spatial paths. $\Delta k(\omega_{as})$ is the phase
mismatching, and $\kappa(\omega_{as})$ is the nonlinear parametric
coupling coefficient. $\kappa(\omega_{as})
\text{sinc}[\frac{\Delta k(\omega_{as})L}{2}]$ determines the
photon spectrum. $\phi$ is the phase difference between the two
SFWM paths. Equation (1) can also be rewritten in the time and
polarization domains
\begin{equation}
\begin{split}
|\Psi(t_{s},t_{as})\rangle& =\psi(t_{as}-t_{s})e^{-i(\overline{\omega}_{as}t_{as}+\overline{\omega}_{s}t_{s})}\\
&
\otimes\frac{1}{\sqrt{2}}(|\sigma_{p}^{1}\rangle_{s}|\sigma_{c}^{1}\rangle_{as}
+e^{i\phi}|\sigma_{p}^{2}\rangle_{s}|\sigma_{c}^{2}\rangle_{as} ),
\end{split}
\end{equation}
where $t_{s}$ and $t_{as}$ are the detection time of the Stokes
and anti-Stokes photons, respectively. $\overline{\omega}_{s}$ and
$\overline{\omega}_{as}$ are their central frequencies,
respectively. The time-domain wave function $\psi(t_{as}-t_{s})$
results from the frequency entanglement
($\omega_s=\omega_{p}+\omega_{c}-\omega_{as}$) and is the Fourier
transform of the two-photon joint spectrum. Meanwhile, as shown in
Eqs. (1) and (2), the polarization entanglement can be manipulated
by controlling the pump-coupling polarizations and phase
difference.

We first characterize the two-photon nonclassical correlation in
the time domain. Figure 2 (a)-2(c) show the two-photon coincidence
counts as functions of the relative time delay ($\tau=t_{as}-t_s$)
with the polarization configurations $\sigma_s^+\sigma_{as}^-$,
$\sigma_s^-\sigma_{as}^+$, and
$\sigma_s^+\sigma_{as}^-+\sigma_s^-\sigma_{as}^+$, respectively.
We carefully balance the pump and coupling laser powers on the two
SFWM paths to make their correlation indistinguishable in the time
domain for achieving the maximally polarization-entangled states.
At each path, the pump  beam has a power of 8 $\mu$W and the
coupling beam of 2 mW. As shown in Fig. 2, these paired photons
have a temporal correlation length of 300ns. Normalizing the
coincidence counts to the accidental uncorrelated background
counts, we obtain the normalized cross correlation
$g_{s,as}^{(2)}(\tau)$ with a peak value of 35 at $\tau$=25 ns.
With measured autocorrelations
$g_{s,s}^{(2)}(0)=g_{as,as}^{(2)}(0)\thickapprox2.0$, we confirm
that the Cauchy-Schwartz inequality
$[g_{s,as}^{(2)}(\tau)]^{2}\leq
g_{s,s}^{(2)}(0)g_{as,as}^{(2)}(0)$ is violated by a factor of
$~306$, which clearly indicates the quantum nature of the paired
photons. With an integration time bin of 300 ns, the normalized
cross correlation has a reduced peak value of 10, which still
violates the Cauchy-Schwartz inequality by a factor of 25. The
solid curves in Fig. 2(a)-2(c) are calculated from Eq. (2). The
nearly perfect agreement between the theory and experiment
indirectly verifies the time-frequency entanglement. The shorter
correlation time in Fig. 2(c) compared to that in Fig.2(a) and (b)
is caused by the addition of the powers from the two coupling
beams that widen the EIT and biphoton bandwidth. The coherence
time of about 300 ns corresponds to an estimated bandwidth of 2.9
MHz (also confirmed from the theory). The photon bandwidth can be
further reduced by lowering the coupling laser power to narrow the
EIT window. Figure 2(d) shows the measured correlation time versus
the coupling laser power. With 0.13 mW coupling laser power, we
obtain a coherence time of up to 900 ns, which corresponds to a
bandwidth of about 0.8 MHz. The solid curve in Fig. 2(d) is also
obtained with Eq. (2).

We next demonstrate that all four polarization-entangled Bell
states can be realized by locking the phase $\phi$ as well as
properly choosing the polarizations of the coupling and pump laser
beams. As shown in Eq. (2), locking the phase difference $\phi$ as
$0$ or $\pi$ and setting $\sigma_p^1=\sigma_c^2=\sigma^+$,
$\sigma_c^1=\sigma_p^2=\sigma^-$, we can produce two
polarization-entangled Bell states:
\begin{equation}
|\Psi^{\pm}\rangle=\frac{1}{\sqrt{2}}\left( {\left| {\sigma _{s}^
+ \sigma _{as}^ -
 } \right\rangle  \pm \left| {\sigma
_{s}^ - \sigma _{as}^ +  } \right\rangle } \right).
\end{equation}
Similarly, by setting $\sigma_p^1=\sigma_c^1=\sigma^+$,
$\sigma_p^2=\sigma_p^2=\sigma^-$, we obtain  other two Bell
states:
\begin{equation} |\Phi^{\pm}\rangle=\frac{1}{\sqrt{2}}\left( {\left|
{\sigma _{s}^ + \sigma _{as}^ + } \right\rangle  \pm \left|
{\sigma _{s}^ - \sigma _{as}^ - } \right\rangle } \right).
\end{equation}

\begin{figure}
\begin{center}
\includegraphics[width=8cm]{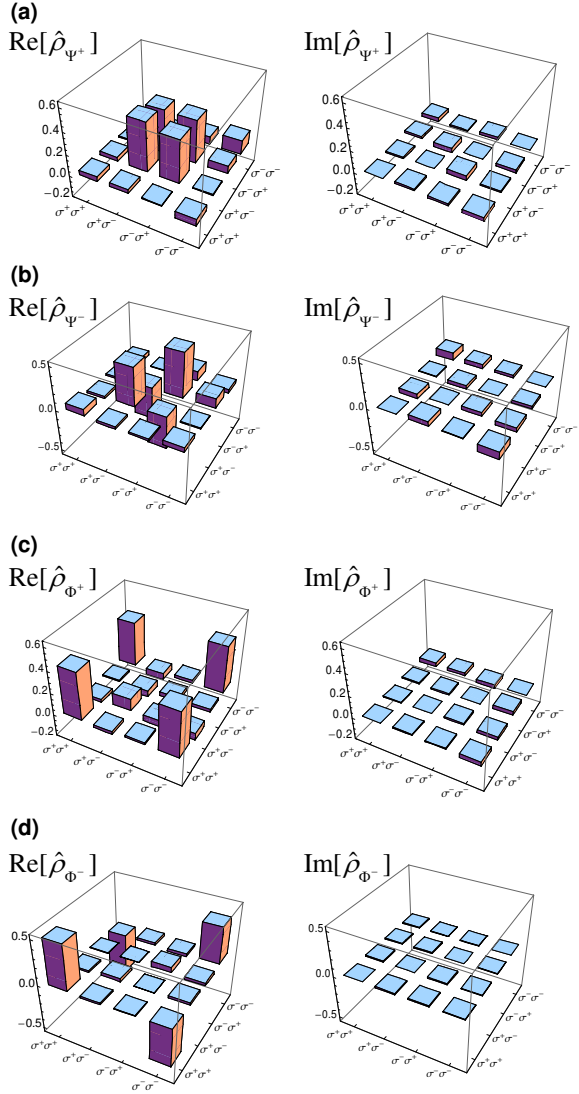}
\caption{\label{fig:4}(color online). Tomography measurement
results of the photon pairs at the target polarization-entangled
Bell states (a) $|\Psi^ +\rangle $, (b) $|\Psi^ -\rangle $, (c)
$|\Phi^ +\rangle $, and (d) $|\Phi^ -\rangle $.}
\end{center}
\end{figure}

Figure 3 displays the measured two-photon polarization
correlations for $|\Psi^{+}\rangle$ by locking $\phi=0$.  The
coincidence counts are integrated from $\tau=0$ to 300 ns for a
total measurement time of 150 s.  The circle data ($\circ$, red
line) are collected by fixing the Stokes photon polarization angle
at $0^\circ$ and the triangle data ($\triangle$, blue line) at
$-45^\circ$. Other parameters during the measurement remain the
same as those for Fig. 2(c). The solid cosine- and sine-wave
curves are the theoretical fits with adjustable background and
amplitude parameters. We obtain the visibility $V = 89.3\%$, which
is beyond the requirement of $1/\sqrt{2}$ for violating the
Bell-Clauser-Horne-Shimony-Holt (Bell-CHSH) inequality\cite{CHSH}.
For comparison, we plot the data without locking the phase as the
square points ($\square$, black line) which shows no quantum
interference.

To obtain a complete characterization of the polarization
entanglement, we also make a quantum state tomography to determine
the density matrix following the maximum likelihood estimation
method \cite{WhitePRL1999, JamesPRA2001}. With two additional
quarter-wave plates, the circular polarization basis $\left|
{\sigma _s^ + \sigma _{as}^ - } \right\rangle$,$\left| {\sigma _s^
- \sigma _{as}^ + } \right\rangle$,$\left| {\sigma _s^ + \sigma
_{as}^ + } \right\rangle$, and $\left| {\sigma _s^ - \sigma _{as}^
- } \right\rangle$ can be converted into linear polarization basis
${\left| HH \right\rangle }$,${\left| VV \right\rangle }$,${\left|
HV \right\rangle }$, and ${\left| VH \right\rangle }$. Then we use
a half-wave plate followed by a PBS as the polarization selector.
The density matrix is constructed from the coincidence counts at
16 independent projection states (See Supplementary Material
\cite{Supp}). The graphical representations of the obtained
density matrix for $|\Psi^ \pm\rangle $ and $|\Phi^ \pm\rangle $
are shown in Fig. 4, from which we obtain the fidelities of
93.6\%, 91.8\% , 92.9\%, and 95.2\%, respectively.  We also use
the density matrix to test the violation of the Bell-CHSH
inequality ($S > 2$) and get the values $S = 2.23 \pm 0.025$,
$2.19 \pm 0.026$, 2.3 $\pm$ 0.02, and 2.39$ \pm$0.026 for the
obtained four Bell states.

\begin{figure}[tb]
\begin{center}
\includegraphics[width=7.5cm]{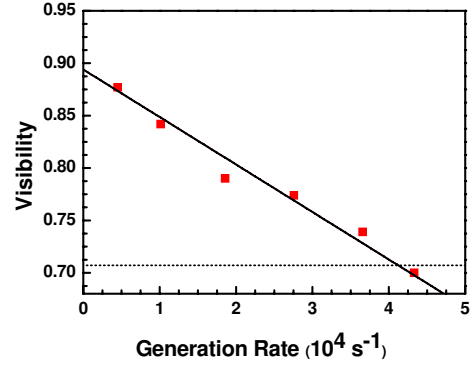}
\caption{\label{fig:5}(color online). Visibility versus the
biphoton generation rate. The solid line is linear fit. The dashed
line marks the boundary for violating the Bell-CHSH inequality. }
\end{center}
\end{figure}

Now, we turn to the brightness of our photon source. For the data
shown in Fig. 2 (c), by taking into account the fiber coupling
efficiency ($70\%$), filter transmission ($70\%$), detector
quantum efficiency ($50\%$), and duty cycle ($10\%$), our photon
source spontaneously generates about $9800$ photon pairs per
second. With the pump power of $16\ \mu \text{W}$ and a linewidth
of $2.9$ MHz, we estimate a spectrum brightness of $3400\
{\text{s}^{ - 1}} \text{MHz}^{ - 1}$ and the normalized spectrum
brightness of $213000\ {\text{s}^{ - 1}} \text{MHz}^{ - 1} \text{
mW}^{-1}$. We can further increase the photon pair generation rate
by increasing the pump laser power. Figure 5 shows the dependence
of the visibility of the polarization correlation to the photon
pair generation rate.  The visibility is estimated from $V =
(g_{s,as}^{(2)} -1)/(g_{s,as}^{(2)} + 1)$ \cite{correlation,
gtwo2}. As long as the generation rate is less than $ 4
\times{10^4}\ {\text{s}^{ - 1}}$ (here the averaged normalized
cross correlation is under consideration), the visibility is
larger than $1/\sqrt{2}$, which is the boundary to violate the
Bell-CHSH inequality.

In summary, we have demonstrated an efficient experimental scheme
for producing subnatural-linewidth photon pairs with polarization
entanglement. The polarization entanglement results from the
interference between the two SFWM spatial paths. By stabilizing
the phase difference between these two paths and setting properly
the driving laser polarizations, we produce all four Bell states,
confirmed by the quantum state tomography measurements. Their long
coherence time (up to 900 ns) and narrow bandwidth (about 1 MHz)
make them a promising entangled photon source for interacting with
rubidium atomic quantum nodes.

We thank J. F. Chen and C. Liu for their helpful discussion. This
work was supported by the NSFC (Grants No. 11104085, No. 11125417,
and No. 61378012), the Major Research Plan of the NSFC (Grant No.
91121023), the SKPBR of China (Grants No. 2011CB922104 and No.
2011CBA00200), the FOYTHEG (Grant No. Yq2013050), the PRNPGZ
(Grant No. 2014010), and the PCSIRT (Grant No. IRT1243). S.D. was
support by Hong Kong Research Grants Council (Project No.
HKU8/CRF/11G).

\newpage

\begin{widetext}

{\bf Supplemental Material:}

\section{Derivation of Eqs. (1) and (2)}
In the interaction picture the effective interaction Hamiltonian
for the four-wave mixing (FWM) parametric process takes the
form\cite{wen1, wen2, review}

\begin{align}
\begin{split}
&\hat{H}_{I}(t)=\frac{i\hbar L}{2\pi}\int
d\omega_{as}d\omega_{s}\kappa \text{sinc}(\frac{\Delta
kL}{2})\frac{1}{\sqrt{2}}[\hat{a}^{+}_{\sigma^{1}_{p}}(\omega_{s})
\hat{a}^{+}_{\sigma^{1}_{c}}(\omega_{as})
+\hat{a}^{+}_{\sigma^{2}_{p}}(\omega_{s})\hat{a}^{+}_{\sigma^{2}_{c}}(\omega_{as})e^{ik_{c}(L_{c2}-L_{c1})}e^{ik_{p}(L_{p2}-L_{p1})}]\\
&\cdot e^{-i(\omega_{p}+\omega_{c}-\omega_{as}-\omega_{s})t}+H.C.,\\
&=\frac{i\hbar L}{2\pi}\int d\omega_{as}d\omega_{s}\kappa
\text{sinc}(\frac{\Delta
kL}{2})\frac{1}{\sqrt{2}}[\hat{a}^{+}_{\sigma^{1}_{p}}(\omega_{s})
\hat{a}^{+}_{\sigma^{1}_{c}}(\omega_{as})
+\hat{a}^{+}_{\sigma^{2}_{p}}(\omega_{s}) \hat{a}^{+}_{\sigma^{2}_{c}}(\omega_{as})e^{i\phi_{1}}e^{i\phi_{2}}]e^{-i(\omega_{p}+\omega_{c}-\omega_{as}-\omega_{s})t}+H.C.\\
\end{split}
\tag{S.1}
\end{align}
where $\hat{a}^{+}_{\sigma^{j}_{p}}(\omega_{s})$ $(j=1,2)$ is the
creation operator of the Stokes photons with pump polarization
$\sigma^j_p$, $\hat{a}^{+}_{\sigma^{j}_{c}}(\omega_{as})$
$(j=1,2)$ is the creation operator of the anti-Stokes photons with
coupling polarization $\sigma^j_c$,
 $L_{c_j}$ ($L_{p_j}$) is the length of the coupling
(pump) laser arm j in the Mach-Zehnder interferometer,
$\phi_{1}=k_{c}(L_{c2}-L_{c1})$, $\phi_{2}=k_{p}(L_{p2}-L_{p1})$.
Based on perturbation theory \cite{review}, the two-photon
(biphoton) state $|\Psi\rangle$ can be expressed as
\begin{align}
\begin{split}
&|\Psi\rangle=-\frac{i}{\hbar}\int^{+\infty}_{-\infty}dt
\hat{H}_{I}(t)|0\rangle\\
&=L\int d\omega_{as} \kappa(\omega_{as}) \text{sinc}(\frac{\Delta
k(\omega_{as})L}{2})\frac{1}{\sqrt{2}}[\hat{a}^{+}_{\sigma^{1}_{p}}(\omega_{p}+\omega_{c}-\omega_{as})\hat{a}^{+}_{\sigma^{1}_{c}}(\omega_{as})
+\hat{a}^{+}_{\sigma^{2}_{p}}(\omega_{p}+\omega_{c}-\omega_{as})\hat{a}^{+}_{\sigma^{2}_{c}}(\omega_{as})e^{i\phi}]|0\rangle\\
&=L\int d\omega_{as} \kappa(\omega_{as}) \text{sinc}(\frac{\Delta
k(\omega_{as})L}{2})\frac{1}{\sqrt{2}}[|\omega_s=\omega_{p}+\omega_{c}-\omega_{as},
\sigma^{1}_{p}\rangle |\omega_{as}, \sigma^{1}_{c}\rangle
+|\omega_s=\omega_{p}+\omega_{c}-\omega_{as},
\sigma^{2}_{p}\rangle |\omega_{as},
\sigma^{2}_{c}\rangle e^{i\phi}]\\
&= L\int d\omega_{as}\kappa(\omega_{as})\text{sinc}[\frac{\Delta
k(\omega_{as})L}{2}]|\omega_s=\omega_{p}+\omega_{c}-\omega_{as}\rangle|\omega_{as}\rangle
\otimes \frac{1}{\sqrt{2}}[
|\sigma_{p}^{1}\rangle_{s}|\sigma_{c}^{1}\rangle_{as}+e^{i\phi}
|\sigma_{p}^{2}\rangle_{s}|\sigma_{c}^{2}\rangle_{as}],\\
\end{split}
\tag{S.2}
\end{align}
where $\phi=\phi_{1}+\phi_{2}$.

Equation (S.2) shows generation of both time-frequency
entanglement (energy conservation due to the time translation
symmetry of the system
$\omega_{s}=\omega_{c}+\omega_{p}-\omega_{as}$) and polarization
entanglement.

For simplify, Equation (S.2) can also be further described in the
time and polarization domain as
\begin{align}
\begin{split}
&|\Psi(t_{s},t_{as})\rangle=\psi(t_{as}-t_{s})e^{-i(\overline{\omega}_{as}t_{as}+\overline{\omega}_{s}t_{s})}\otimes\frac{1}{\sqrt{2}}[
|\sigma_{p}^{1}\rangle_{s}|\sigma_{c}^{1}\rangle_{as}+e^{i\phi}
|\sigma_{p}^{2}\rangle_{s}|\sigma_{c}^{2}\rangle_{as}].
\end{split}
\tag{S.3}
\end{align}

In Eq. (S.3), the time-domain wave function $\psi(t_{as}-t_{s})$,
which indicates the time-frequency entanglement, is the Fourier
transform of the photon spectrum
$\kappa(\omega_{as})\text{sinc}[\frac{\Delta k(\omega_{as})L}{2}]$
and can be measured through coincidence counts in the time domain
directly. Equations (S.2)  and (S.3) correspond to Eqs. (1) and
(2) in the main text, respectively.

\section{Method to lock phase factor $\phi$}

In order to produce the polarization entangled Bell state, the
phase difference $\phi$ in Eq.(2) in the main text should be
locked. For the experimental setup, the phase difference is given
by
\begin{align}
\phi=\frac{2\pi}{\lambda_{c}}(L_{c2}-L_{c1})+\frac{2\pi}{\lambda_{p}}(L_{p2}-L_{p1}),
\tag{S.4}
\end{align}
where $\lambda_c$ ($\lambda_p$) is the wave length of the coupling
(pump) laser beam. Let's define
\begin{align}
k_{0}=\frac{k_{c}+k_{p}}{2}, \delta=\frac{k_{c}-k_{p}}{2},
\tag{S.5}
\end{align}
then the phase difference can be rewritten as
\begin{align}
\phi=k_{0}[(L_{c2}+L_{p2})-(L_{c1}+L_{p1})]+\delta
[(L_{c2}-L_{p2})-(L_{c1}-L_{p1})]. \tag{S.6}
\end{align}

Under the condition $k_{0}\gg \delta$, Equation (S.6) can be
further reduced to
\begin{align}
\phi\approx
k_{0}[(L_{c2}+L_{p2})-(L_{c1}+L_{p1})]=\frac{k_{0}}{k_{l}}\varphi+\varphi_{0},
\tag{S.7}
\end{align}
here $\varphi=k_{l}[(L_{c2}+L_{p2})-(L_{c1}+L_{p1})]$ is the phase
difference of the Mach-Zehnder interferometer for the locking
laser ($k_{l}$, 795nm), $\varphi_{0}$ is the phase shift between
the two Mach-Zehnder interferometers (the locking laser doesn't
overlap with the pump-coupling laser fields).

We use the two-photon polarization correlation measurement to
determine the relation between the locking point $\varphi$ and the
phase difference $\phi$. As shown in Fig. (S.1), with four groups
of correlation measurements ($0,\pi/3, 2\pi/3$ and $\pi$), the
locking point $\varphi$ are found to be linearly relative to the
phase difference $\phi$ that fitted very well with the theoretical
results from Eq. (S.7).

\setcounter{figure}{0}
\renewcommand{\thefigure}{S\arabic{figure}}
\begin{figure}
\begin{center}
\includegraphics[width=8.5cm]{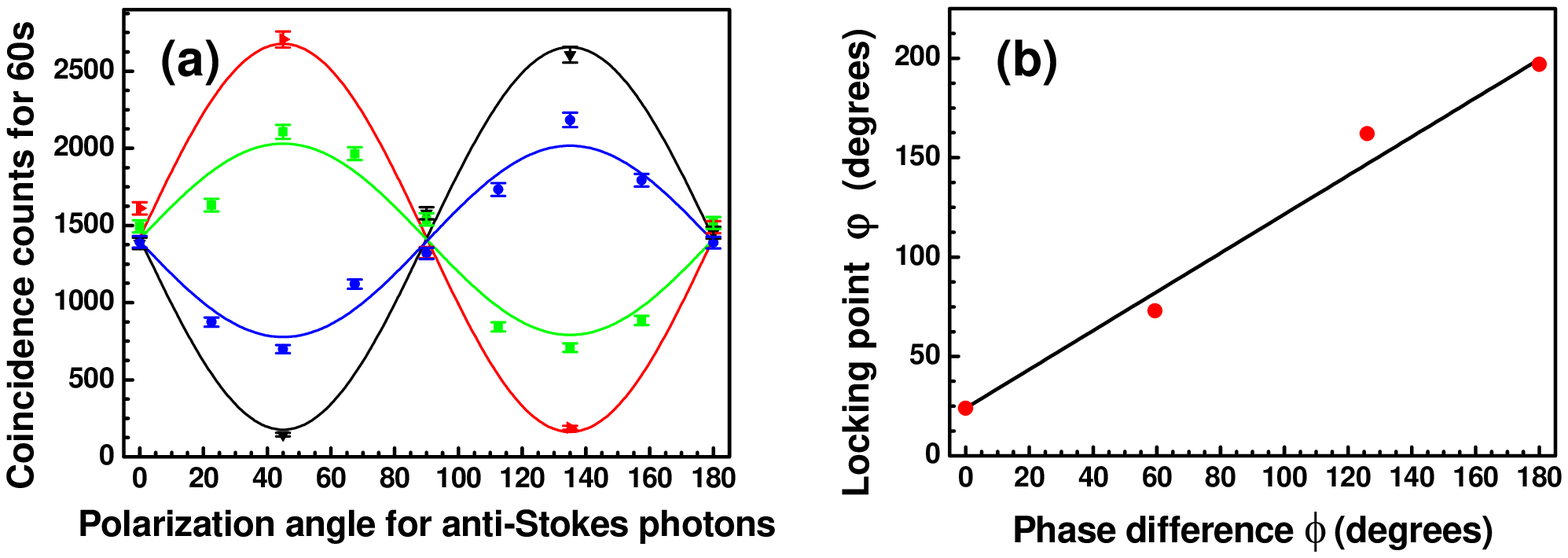}
\caption{\label{fig:1} (color online). (a) Polarization
correlation of the paired Stokes and anti-Stokes photons. The
polarization angle for Stokes photons are set at $-45^\circ$, the
phase difference $\phi$ is $0$ (black line and cycle), $\pi/3$
(red line and triangle), $2\pi/3$ (blue line and star) and $\pi$
(green line and square), respectively.  (b)The locking point
versus the phase difference of the polarization entangled state.
All of the solid lines are theoretically fitted.}
\end{center}
\end{figure}

\section{Density matrix results of the four Bell states}
The properties of the four Bell states generated in our
experiments are described in the main text. The detailed density
matrix results are showing in the following:

\[ |\Psi^{+}\rangle:\ \ \ \left( {\begin{array}{*{20}{c}}
{0.042}&{0.034 - 0.014i}&{ 0.008 + 0.018i}&{ - 0.06 + 0.032i}\\
{ 0.034 + 0.014i}&{ 0.449}&{0.393 - 0.04i}&{ 0.008 - 0.011i}\\
{ 0.008 - 0.018i}&{ 0.393 + 0.04i}&{ 0.387}&{ 0.069 + 0.025i}\\
{ - 0.06 - 0.032i}&{0.008 + 0.011i}&{ 0.069 - 0.025i}&{ 0.123}
\end{array}} \right),\]

\[ |\Psi^{-}\rangle:\ \ \ \left( {\begin{array}{*{20}{c}}
{ 0.071}&{ - 0.024 - 0.05i}&{ - 0.019 - 0.006i}&{ 0.047 + 0.082i}\\
{ - 0.024 + 0.05i}&{ 0.449}&{ - 0.386 - 0.037i}&{  0.009 + 0.045i}\\
{ - 0.019 + 0.006i}&{ - 0.386 + 0.037i}&{ 0.52}&{ 0.109 + 0.027i}\\
{0.047 - 0.082i}&{  0.009 - 0.045i}&{ 0.109 - 0.027i}&{-0.041}
\end{array}} \right),\]

\[ |\Phi^{+}\rangle:\ \ \ \left( {\begin{array}{*{20}{c}}
{ 0.432}&{ -0.024 - 0.008i}&{ 0.013 + 0.009i}&{ 0.407+0.037i}\\
{ -0.024 + 0.008i}&{ 0.104}&{ - 0.055 -0.014i}&{- 0.06 + 0.025i}\\
{ 0.013 - 0.009i}&{ -0.055 + 0.014i}&{ 0.034}&{ 0.025 - 0.031i}\\
{0.407 - 0.037i}&{- 0.006 -0.025i}&{ 0.025 + 0.031i}&{0.431}
\end{array}} \right),\]

\[ |\Phi^{-}\rangle:\ \ \ \left( {\begin{array}{*{20}{c}}
{ 0.5}&{ - 0.018 + 0.012i}&{ - 0.006 + 0.009i}&{ -0.438 + 0.008i}\\
{ - 0.018 - 0.012i}&{ 0.034}&{ 0.001 + 0.013i}&{0.024 + 0.005i}\\
{ -0.006 - 0.009i}&{ 0.001 - 0.013i}&{ 0.056}&{ - 0.028 - 0.006i}\\
{-0.438 - 0.008i}&{0.024 -0.005i}&{ - 0.028 + 0.006i}&{0.409}
\end{array}} \right).\]

\end{widetext}

\end{document}